\documentclass[11pt,leqno]{article}

\usepackage{mathcomp}
\usepackage{graphicx}
\usepackage{graphics}
\usepackage{epstopdf}
\usepackage{epsfig}
\usepackage{amsthm,mathrsfs}
\usepackage{amsmath}
\usepackage{srcltx}

\DeclareSymbolFont{AMSb}{U}{msb}{m}{n}
\DeclareSymbolFontAlphabet{\mathbb}{AMSb}

 \newcommand{\beqn}{\begin{eqnarray}}
 \newcommand{\eeqn}{\end{eqnarray}}
 \newcommand{\be}{\begin{equation}}
 \newcommand{\ee}{\end{equation}}
 \newcommand{\ba}{\begin{array}}
 \newcommand{\ea}{\end{array}}

 \newcommand{\re}{\ref}
 \newcommand{\ci}{\cite}
 \newcommand{\ds}{\displaystyle}%%%display%%%
 \newcommand{\la}{\label}
 \newcommand{\bfr}{\begin{flushright}}
 \newcommand{\efr}{\end{flushright}}

\newcommand{\bfl}{\begin{flushleft}}
\newcommand{\efl}{\end{flushleft}}

\newcommand{\pr}{\prime}

\def\Re {{\rm Re\, }}                                       % Re
\newcommand{\ov}{\overline}

\newcommand{\loota}{\hbox{\enspace{\vrule height 7pt depth 0pt width
      7pt}}}

\newcommand{\bo}{{\hfill\loota}}

\newcommand{\E}{{\cal E}}
\newcommand{\bF}{{\bf F}}

\newcommand{\ddd}{\stackrel{.\kern-.07em.\kern-.07em.}}
\newcommand{\bre}{|\kern-.25em|\kern-.25em|}
 \def\R{{\rm I\kern-.1567em R}}                              % Doppel R
 \def\C{{\rm C\kern-4.7pt                                    % Doppel C
 \vrule height 7.5pt width 0.4pt depth -0.3pt \phantom {.}}}

 \def\Z{{\sf Z\kern-4.5pt Z}}                                % Doppel Z

 \def\N{{\rm I\kern-.1567em N}}                              % Doppel-N

\newtheorem{theorem}{Theorem}[section]
\newtheorem{qtheorem}{QTheorem}[section]

\newtheorem{defin}[theorem]{Definition}

\newtheorem{lemma}[theorem]{Lemma}
\newtheorem{example}[theorem]{Example}
\newtheorem{examples}[theorem]{Examples}
\newtheorem{exercice}[theorem]{Exercise}
\newtheorem{remark}[theorem]{Remark}
\newtheorem{remarks}[theorem]{Remarks}
\newtheorem{cor}[theorem]{Corollary}
\newtheorem{pro}[theorem]{Proposition}

\newcommand{\bd}{\begin{defin}}
 \newcommand{\ed}{\end{defin}}
\newcommand{\bt}{\begin{theorem}}
 \newcommand{\et}{\end{theorem}}
\newcommand{\bqt}{\begin{qtheorem}}
 \newcommand{\eqt}{\end{qtheorem}}

\newcommand{\bp}{\begin{pro}}
 \newcommand{\ep}{\end{pro}}

\newcommand{\bl}{\begin{lemma}}
 \newcommand{\el}{\end{lemma}}
\newcommand{\bc}{\begin{cor}}
 \newcommand{\ec}{\end{cor}}

\newcommand{\bex}{\begin{example}}
 \newcommand{\eex}{\end{example}}
\newcommand{\bexs}{\begin{examples}}
 \newcommand{\eexs}{\end{examples}}

\newcommand{\bexe}{\begin{exercice}}
 \newcommand{\eexe}{\end{exercice}}

\newcommand{\br}{\begin{remark} }
 \newcommand{\er}{\end{remark}}
\newcommand{\brs}{\begin{remarks}}
 \newcommand{\ers}{\end{remarks}}

\begin{document}

\begin{center}

\vspace{3cm}

{\huge\bf On Asymptotic Completeness of Scattering
  \bigskip\\
in the Nonlinear Lamb\bigskip\\ System, II}
\\
\vspace{1cm}

{\large A.I.Komech \footnote{
Supported partly by Alexander von Humboldt Research Award, Austrian
Science Fund (FWF): P22198-N13, and the grants of DFG and RFBR,
}}\\
%%%%%%%%%%%%%%%

%%%%%%%%%%%%%%%%%

{\it
Faculty of Mathematics of Vienna University\\
and the Institute for the Information Transmission Problems of
RAS (Moscow)}\\
e-mail:~alexander.komech@univie.ac.at
 \bigskip\\
 {\large A.E.~Merzon
\footnote{Supported  by CONACYT and CIC of
 UMSNH and FWF-project P19138-N13.}}\\

{\it Institute of Physics and  Mathematics\\
University of Michoac\'{a}n of
 San Nicolas de Hidalgo \\
 Morelia, Michoac\'{a}n, M\'{e}xico }\\
  e-mail:~anatoli@ifm.umich.mx.
\bigskip\\

\end{center}

\begin{abstract}
We establish the asymptotic completeness in the nonlinear
Lamb system for hyperbolic stationary states. For the proof we construct a trajectory of a
reduced equation (which is a nonlinear nonautonomous
ODE) converging to a hyperbolic stationary point using the Inverse Function Theorem in a Banach space. We give the counterexamples showing nonexistence of such trajectories for nonhyperbolic stationary points.

\end{abstract}

%%%%  NACHAlO TEXTA    %%%%%%%%%%%%%%%%%%%%%%%%%%%%%%%%%%%%%
%%%%%%%%%%%%%%%%%%%%%%%%%%%%%%%%%%%%%%%%%
%%%%%%%%%%%%%%%%%%%ix  %%%%%%%%%%%%%%%%%%%%%%

 \newpage\hspace{-7mm}

\section{Introduction}
\setcounter{equation}{0}

In this paper we consider the asymptotic completeness
of scattering in the
nonlinear Lamb
system for the case of
zero oscillator mass $m=0$.
This system describes the string coupled to the n-dimensional nonlinear oscillator
with the force function
$F(y)$, $y\in\R^n$,
\begin{eqnarray}\la{L1.1}
\left\{ \ba{rcl} \ddot u(x,t)&=&u^{\pr\pr}(x,t),
~~~x\in\R\setminus\{0\},\medskip\\
0&=&F(y(t))+u^\pr(0+,\,t)-u^\pr(0-,\,t);\,~~y(t):=
u(0,t), \ea\right.
\end{eqnarray}
where $\ds\dot u:=\frac{\partial u}{\partial t}$,
$\ds u':=\frac{\partial u}{\partial x}$ . The solution $u(x,t)$ takes
the values in $\R^n$ with $n\geq 1$.

The system (\re{L1.1}) has been
introduced
originally by H. Lamb \cite{Lam}
in the linear case when
$F(y)=-\omega^2 y$ and $n=1$.
The Lamb system with general nonlinear function $F(y)$ and
the oscillator of mass
$m\ge 0$
has been considered in \ci{KB}
where the questions of
irreversibility and nonrecurrence were discussed.
The system was
studied further
in \ci{KMo} where the global attraction
to stationary states has been established
for the first time,
and in \ci{FK06}
where metastable regimes were studied for the stochastic
Lamb system.
The scattering asymptotics with a diverging free wave were established in \cite {km9}.

%%%%%%%%%% 55555555555555
In present paper we continue the study of the
asymptotic completeness in the nonlinear scattering for the Lamb system.
The case $n=1$ was studied in \cite {km}
for hyperbolic stationary states
under condition $F'(0)\not = 0.$
We prove the asymptotic completeness
for all $n>1$ in the hyperbolic case.

The  asymptotics completeness for nonlinear
wave equations was considered in
\cite{S81}) for small initial  states.
We prove the asymptotic completeness
without the smallness assumption.

The paper is organized as follows. In Section 2 we introduce
basic notations, and we
recall
some
statements and constructions from
\cite {KMo,Ka05, km, mt}.
In Section 3 we reduce the asymptotic completeness to the existence of incoming trajectory
of a reduced ODE.
In Section 4 we prove the existence of the incoming trajectory
for small perturbations. First, we prove this for linear $F$
and then for nonlinear $F$ using the Inversion Function Theorem.
In Section 5 we extend the results of Section 4
to arbitrary perturbations without  the smallness assumption.
First, the solution is constructed for
large $t$ and then is continued back using a priory estimate.

\section{Scattering asymptotics for the Lamb system}
\setcounter{equation}{0}

We consider
the Cauchy problem for the system (\re{L1.1}) with the
initial conditions
\begin{equation}\la{L2.1}
u|_{t=0}=u_0(x);\,\dot u|_{t=0}=v_0(x).
\end{equation}
 %%%%%%%%%%%%%%%%%%%%%%%%%%%%%%%%%%%%%%%%%%%
Denote by $\Vert\cdot\Vert_{L^2}$ the norm in
the Hilbert space
 $L^2(\R,\R^n)$.
\begin{defin}\la{Ld2.2} The phase space ${\cal E}$
 of finite energy states for the system (\re{L1.1})
is the Hilbert space of the pars
 $(u(x),v(x))\in H^1(\R,\R^n)$ $\oplus L^2(\R,\R^n)$
 with $u^\prime (x)\in L^2(\R,\R^n)$ and the global energy norm
\begin{eqnarray*}%\la{L2.2}
\begin{array}{l}
\Vert (u,v)\Vert_{\cal E}=\Vert u^\prime \Vert+|u(0)|+ \Vert
v\Vert.\\
\end{array}
\end{eqnarray*}
\end{defin}
The Cauchy problem (\re{L1.1}), (\re{L2.1}) can be written in the form
\begin{equation}\la{CP}
\dot Y(t)=\bF(Y(t)),~~~~t\in\R;~~Y(0)=Y_0
\end{equation}
where $Y(t):=(u(\cdot,t),\dot u(\cdot,t))$, and
$Y_0=(u_0,v_0)$ is the initial date.
In \cite{Ka05,mt},
the scattering asymptotics have been proved,
\be\la{scata}
Y(t)\sim S_\pm+W(t)\Psi_\pm,~~~~~~~t\to\pm\infty
\ee
where $S_\pm$ are the limit stationary states,
$W(t)$ is the dynamical group of the free wave equation, and
$\Psi_\pm\in\E$ are the corresponding {\it asymptotic states.}

The asymptotics (\re{scata}) hold in the norm of the Hilbert phase
space $\E$ if the following limits exist:
\begin{equation}\la{limi}
u_0^+:=\lim_{x\to+\infty}u_0(x),~~
u_0^-:=\lim_{x\to-\infty}u_0(x),~~~ \ov v_0:=\int_{-\infty}^\infty
v_0(y)dy.
\end{equation}

%%%%%%%%%%%%%%%%%  11111111111111
We denote by $\cal E_\infty$ the subspace of $\cal E$ consisting of functions  satisfying (\ref{limi}).

%%%%%%%%%%%%%%%%%%%%%%%%2222222222222222
The stationary states $S(x)=(s(x),0)\in
{\cal E}$
for (\re{CP}) are evidently determined with $s(x)\equiv s\in Z=\{z\in\R^n:~F(z)=0\}$.
We denote by ${\cal S}$  the set of all stationary states of system (\ref{L1.1}).
%%%%%%%%%%%%%%%%%%%%%%%%%%%%%%%%%%%%%%%%%%%%%%%%%%%%33333333333333333
The following theorem is proved in \ci[Theorem 4.5 ii) b)]{Ka05} and \ci{mt}, Theorem 3.1.
We assume that
\beqn
F(u)=-\nabla V(u),
~~~ V(u)\in C^2(\R^n,\R),~~\mbox{and}~~
%%\la{L1.2a}
~~~V(u)\to+\infty,~~|u|\to\infty.
\la{L1.2b}
\eeqn
%%%%%%%%%%%%%%%%%%%%%%%%%%%%%%%%%%%%%%%%%%%%%%%%%%%
\begin{pro}\la{Lt3.1}
Let the assumptions (\re{L1.2b}) and (\re{limi}) hold,
$Z$ be a discrete subset in $\R^n$, and
initial state $Y_0\in\E_\infty $.
 Then
\\
i)
For the corresponding solution $Y(t)\in C(\R, \E)$ to the
Cauchy problem
 (\re{CP}),
the scattering asymptotics hold
\begin{equation}\la{scaL}
Y(t)=S_++W(t)\Psi_++r_+(t),~~~~~~t\ge 0,
\end{equation}
with some limit stationary state $S_+\in\cal S$ and
asymptotic state $\Psi_+\in\E_\infty$. The remainder is
small in the global energy norm:
 \begin{equation*}%\la{remL}
 \Vert
r_+(t)\Vert_\E\to 0,\,\,\,t\to\infty.
\end{equation*}
\end{pro}
%%%%%%%%%%%%%%%%%%%%%%%%%%%%%%%%%%444444444444444444
We will call $(S_+,\Psi_+)$ as the scattering data
of the solution $Y(t).$
Our goal is to describe all {\it admissible pairs} $(S_+,\Psi_+)\in \cal S\times \cal E_\infty$ such that there exists $Y_0\in {\cal E}_{\infty}$ satisfying (\ref{scaL}).

Let us comment on previous results in this direction (see \cite {km}, Lemma 2.7, Lemma 5.1 and Theorem 6.1.)
\medskip
\\
{\bf A.} Any asymptotic state $\Psi_+=(\Psi_0,\Psi_1)\in {\cal E}\subset {\cal E}_{\infty}$ satisfies
the identity
 \begin{equation}\la{psiunoi}
\Psi_0^+  + \Psi_0^-+\ov\Psi_1=0
 \end{equation}
where
$ \Psi_0^+=\lim\limits_{x\to+\infty} \Psi_0(x)$,
$\Psi_0^-=\lim\limits_{x\to-\infty}\Psi_0(x)$, and
$\ov{\Psi}_1=\ds\int_{-\infty}^{\infty}\Psi_1(y)dy$.
\medskip\\

We denote by $\cal E_{\infty}^+$ the subspace of ${\cal E}_{\infty}$ consisting
of functions satisfying (\ref{psiunoi}).
\medskip
\\
{\bf B.} A pair $(S_+,\Psi_+)$ is admissible if
$\Psi_+(x)\in\E_\infty ^+$
satisfies the identity (\re{psiunoi}) and
has a compact support.
\medskip
\\
{\bf C.} For  $n=1$
any pair $(S_+,\Psi_+)\in {\cal S}\times \cal E_{\infty}^+ $ with $S_+=(s_+,0)$
is admissible if $F'(s_+)\not =0$.
\medskip
\\
The similar results hold when $t\to-\infty$. Then
the asymptotics (\ref{scaL}) take the form
\begin{equation*}%\la{scaL-}
Y(t)=S_-+W(t)\Psi_-+r_-(t),~~~~~~t\leq 0,
\end{equation*}
where for $\Psi_-=(\Psi_0,\Psi_1).$ The relation (\ref{psiunoi}) is
changed to the following
\begin{equation*}%\la{psiunoi-}
\Psi_0^+  + \Psi_0^--\ov\Psi_1=0.
\end{equation*}
In this paper we generalize the result {\bf C}
for an arbitrary $n>1$ for hyperbolic stationary states.
Let $\sigma(A)$ denote the spectrum of an $n\times n$ -matrix $A$.
\begin{defin}\la{dsd}
The stationary state $S_+=(s_+,0)$ of  system (\ref{L1.1}) is hyperbolic if $\Re \lambda \not = 0$ for all
$\lambda \in \sigma(F'(s_+))$.
\end{defin}
We will prove that a pair $(S_+,\Psi_+)$ is admissible for $n>1$ in the case
of hyperbolic stationary state $S_+$,
and arbitrary $\Psi_+\in \cal E^+_{\infty}$.

\section{Reduced equation}
\setcounter{equation}{0}
Let $Y(t)\in C(\R,\E)$
be a solution to (\re{CP})
with
$Y_0 \in \E_\infty$.
Let us set
\begin{equation}\la{sd}
W_+Y_0=(S_+,\Psi_+)\in {\cal S}\times \E_{\infty}^+
\end{equation}
where $\Psi_+$ is defined by (\ref{scaL}), and
$S_+=(s_+,0)$.
\begin{defin}\la{ascom}
The Lamb system (\ref{L1.1}) is asymptotically complete at a stationary state $S_+$ if
for any $\Psi_+\in {\cal E} _{\infty}^+$
there exists initial data $Y_0\in {\cal E}_\infty$ such that (\ref{sd}) holds.
\end{defin}
For $\Psi_+=(\Psi_0,\Psi_1)\in {\cal E}_\infty^+,$ let us
set
\be\la{S}
 S(t):=\frac{\Psi_0(t) + \Psi_0(-t)}{2} +
\frac{1}{2}\int_{-t}^t \Psi_1(y)dy,~~ t\in\R. \ee

\bl\label{ty0}(\cite{km}, Lemma 3.1) Let $Y(t)\in C(\R,\E)$ be a solution of (\ref{CP})
with $Y(0)=Y_0\in {\cal E}_{\infty}$, and (\ref{sd}) holds for some $S_+$ and $\Psi_{+}\in {\cal E}_{\infty}^+$.
Then
$\dot S\in L^2(\R,\R^n)$, and
\be\la{re2}
\dot{y}(t)= -\frac{1}{2}F(y(t))+\dot{S}(t),~~~t > 0 ~;~~
\dot y\in L^2(\R_+,\R^n)~;~~y(t)\to s_+,~t\to
+\infty
\ee
where $y(t):=u(0,t)$.
\el
We call the differential equation  (\ref {re2}) the {\it inverse reduced equation}.
It plays the crucial role in the proof of the asymptotics completeness.
The following lemma, proved in \cite {km}, reduces the problem of asymptotic completeness to the construction of solutions to (\ref{re2}).
\bl\label{tisp} (see \cite {km}, Lemma 4.1)
Let $(S_+,\Psi_+)\in {\cal S}\times {\cal E}_\infty^+$, and
there exists a solution $y(t)$ to
(\ref{re2}). Then there exists $Y_0\in {\cal E}_{\infty}$ such that (\ref{sd}) holds.
\el
Lemma \ref{ty0} implies the existence of solution to (\ref{re2}) for any $\Psi_+\in {\cal E}_\infty^+$ if the system (\ref{CP}) is asymptotically complete at $S_+$. Conversely, Lemma \ref{tisp} implies the asymptotic completeness when (\ref{re2}) has a solution for any $\Psi_+\in {\cal E}_{\infty}^+$.
Thus, (\ref{re2}) gives a characterization of admissible asymptotic states.
%%%%%%%%%%%%%%%%%%%%%%%%%%%%%%%%%%%%%%%%%%%%%%%%%%%%%%%%%%%%%%%%%%%%%%%%%%%%%%%%%%

\section{Incoming trajectories}
\setcounter{equation}{0}
In this section we prove
the existence of a solution to (\ref{re2})
for small $\|\dot S\|_{L^2}$ in the case of
hyperbolic stationary  state $S_+=(s_+,0).$
We adapt to our case the methods \cite{bv}, \cite{vz}
of construction of stable and unstable invariant manifolds in the hyperbolic case.
Namely, first we prove the existence for the linear $F(y)$ and then for the nonlinear $F(y)$
 with small perturbations. We will extend these results to arbitrary perturbations $\dot S\in L^2(\R,\R^n)$ in the next section.

\subsection{Linear equation}
\setcounter{equation}{0}

Let $A$ be a linear operator $\R^n\to \R^n$
and
$$
\Re \lambda\not = 0,~~\lambda\in \sigma(A).
$$
Then
$$
{\rm Sp}~ A=\sigma_-\cup \sigma_+,
$$
where $\Re \lambda<0$ for all $\lambda \in \sigma_-$ and $\Re \lambda>0$ for all $\lambda \in \sigma_+.$
Let $\varepsilon>0$ be such that
\begin{equation}\label {lam}
|\Re \lambda|>\varepsilon,~~\lambda\in \sigma(A).
\end{equation}
Denote by $P_{\pm}$ the projectors of $\R^n$ to the subspaces
generated by the eigenvectors corresponding to $\sigma_\pm$ respectively.
Then the operator $A$ is decomposed
$$
A=A_+ + A_-,~~A_{\pm}=AP_{\pm}.
$$
\bd\label{df}
Define the Banach space
$${\cal Y}:=L^2\cap C_b^0$$
where
\begin{equation*}%\label{f}
L^2:=L^2(\R,\R^n)~, C_b^0:=\{y\in C_b(\R,\R^n):~y(t)\to 0,~~t\to \infty\},
\end{equation*}
and  for $y\in {\cal Y}$
$$
\|y\|:=\|y\|_{L^2}+\|y\|_{C_b}.
$$
\ed
Consider
\begin{equation}\label{ls}
\dot y(t)=Ay(t)+f(t),~~t\in \R.
\end{equation}
\begin{lemma}\label{1.1}
 There exists a continuous linear operator $R:L^2\to {\cal Y}$
such that for any $f\in L^2$ and $y\in {\cal Y}$
equation (\ref{ls}) is equivalent to  $$y=Rf.$$
\end{lemma}
\hspace{-5mm}{\bf Proof.}
Let us introduce a fundamental solution of system (\ref{ls})
\begin{equation*}%\label{et}
E(t):=\left \{
\begin{array}{ll}
e^{A_-t},&t>0,\\
-e^{A_+t},&t<0.
\end{array}
\right.
\end{equation*}
By
(\ref{lam}) we have
\begin{equation}\label{cotaet}
|E(t)|\leq Ce^{-\varepsilon |t|},~~ t\in \R.
\end{equation}
%%%%%%%%%%%%%%%%%%%%%%%%%
%%%%%%%%%%%%%%%%%%%%%%%%%%%%%%%%%%%%%%%%%%%%%%%%%%%%%%%%%%
Let us check
that
\begin{equation}\label{zpm}
y =Rf:=E\ast f.
\end{equation}
is a solution to (\ref{ls}), belongs to $\cal Y$ and tends to $0$, as $t\to \infty$.
\medskip
\\
i) Obviously, $y$ satisfies (\ref{ls}). Now let us prove that $y\in C_b$.
By (\ref{cotaet}) we have:
\begin{equation}\label{estcon}
\begin{array}{ll}
|y(t)|&=|(E\ast f)(t)|\leq C\int\limits_{-\infty}^{\infty} e^{-\varepsilon|t-s|}|f(s)|ds
\leq C\|f\|_{L^2}
\end{array}
\end{equation}
by the Cauchy-Schwartz inequality.
\medskip
\\
ii) Let us check that $y\in L^2$. Denote $M(t):=|f(t)|$.
Passing to the Fourier transform $M\to \tilde M$ and using  (\ref{cotaet}) and (\ref{zpm}) we obtain
$$
\|y\|_{L^2}\leq \|e^{-\varepsilon |t|}\ast M(t)\|_{L^2} =\frac{1}{2\pi}\|\frac{2\varepsilon}{\omega^2+\varepsilon^2}\tilde M(\omega)\|_{L^2}
\leq C\|\tilde M\|_{L^2}=2\pi C\|f\|_{L^2}.
$$
iii) Finally, let us prove that $y(t)\to 0$, as $t\to \infty$.
By (\ref{estcon})  it suffices to check that
$$
\int\limits_{-\infty}^{t/2} e^{-\varepsilon|t-s|}|f(s)|ds\to 0,~~\int\limits_{t/2}^{\infty} e^{-\varepsilon |t-s|}|f(s)|ds\to 0,~~t\to\infty.
$$
The second limit follows  from the Cauchy-Schwartz inequality since
$$
\|f\|_{L^2(t/2,\infty)}\to 0,~~t\to \infty.
$$
It remains to prove the first limit. The limit holds since
$$
\int\limits_{-\infty}^{t/2} e^{-\varepsilon|t-s|}|f(s)|ds
\leq C e^{-\varepsilon t/2}
\|f\|_{L^2}\to 0,~~t\to\infty. $$
 \bo

\subsection{Nonlinear equation: Inverse Function Theorem}

Let us consider the nonlinear equation:
\begin{equation}\label{de}
\dot y=Ay+N(y)+f(t),~~~~t>0
\end{equation}
where $f\in L^2,$ and
\begin{equation}\label{N}
N\in C^2(\R^n,\R^n),~~ N(0)=0~ {\rm and}~  \nabla N(0)=0.
\end{equation}
The  function $N$ may be considered as the
functional map:
$$
({\cal N}(y))(t):=N(y(t)),~~t\in \R.
$$
\begin{lemma}\label{ln}
i) The map ${\cal N}: {\cal Y}\to {\cal Y}$ is continuous.
\medskip
\\
ii) There exists the Frechet derivative
${\cal N}'(y)\in {\cal L}({\cal Y},{\cal Y})$ for $y\in {\cal Y}.$
\medskip
\\
iii) Moreover,
$$%\begin{equation}\label{N'}
N'(0)=0.
$$
\medskip
\\
iv) $$N'\in C({\cal Y},{\mathscr{L}}({\cal Y},{\cal Y})).$$
\end{lemma}
\hspace{-5mm}{\bf Proof.} i) Conditions
(\ref{N}) imply that for any $\delta>0$
\be\label{ocn}
|N(y)|\leq C_\delta|y|,~~ |y|<\delta.
\ee
Hence ${\cal N}(y)\in {\cal Y}$ for $y\in {\cal Y}$. The Lagrange formula implies that the
map ${\cal N}$ is continuous from ${\cal Y}$ to ${\cal Y}$  by (\ref{N}).
\medskip
\\
ii)
By (\ref{N}) we have
$$
N(y)-N(y_0)=N'(y_0)(y-y_0)+r(y,y_0), ~~
|r(y,y_0)|\leq \alpha(|y-y_0|)|y-y_0|,~~|y|,|y_0|\leq \delta.
$$
for any $\delta>0$, where $\alpha (s)$ is a monotone increasing function of $s\geq 0$ and
\begin{equation*}%\label{M}
\alpha(s)\to 0,~~s\to 0.
\end{equation*}
Hence,
$$
|r(y(t),y_0(t))|\leq \alpha(|y(t)-y_0(t)|)(|y(t)-y_0(t)|).
$$
Therefore,
$$
\|r\|_{C_b}\leq\alpha(\|y-y_0\|_{C_b})\|y-y_0\|_{C_b},~~\|r\|_{L^2}\leq \alpha(\|y-y_0\|_{C_b})\|y-y_0\|_{L^2}.
$$
Now
$$
{\cal N}(y)-{\cal N}(y_0)={\cal N}'(y_0)(y-y_0))+r(y,y_0),~~\|r\|\leq \alpha(\|y-y_0\|)\|y-y_0\|,~~\|y\|,\|y_0\|\leq \delta.
$$
\medskip
\\
iii)
Let us check that ${\cal N}'=0$. By (\ref{N}) it suffices to prove that
$$
\frac{\|{\cal N}(y)\|}{\|(y)\|}\to 0,~~\|y\|\to 0.
$$
This follows from the estimate
$$
|N(y)|\leq C_{\delta} |y|^2,~~|y|<\delta
$$
which holds by (\ref{N}).
\medskip
\\
iv)
We should prove that the map
$$
y\to {\cal N}'(y)
$$
is continuous: ${\cal Y}\to {\cal} L({\cal Y},{\cal Y})$, i.e.
\begin{equation}\label{A}
\|{\cal N}'(y_1)-{\cal N}'(y_2)\|_{{\cal L}({\cal Y},{\cal Y})}\to 0,~~\|y_1-y_2\|\to 0.
\end{equation}
Indeed, (\ref{A}) means that
\begin{equation}\label{A'}
\sup\limits_{\|y\|\leq 1}
\|[{\cal N}'(y_1)-{\cal N}'(y_2)]y\|\to 0,~~\|y_1-y_2\|\to 0.
\end{equation}
Denote by
${\cal N}_{ij}'(y):=\frac{\partial}{\partial y_i}N_j(y)$. Then (\ref {A'}) is equivalent to
\begin{equation*}%\label{b}
\|\frac{\partial}{\partial y_i}N_j(y_1(t))-\frac{\partial}{\partial y_i}N_j(y_2(t))\|\to 0,~~\|y_1-y_2\|\to 0.
\end{equation*}
Finally, this follows from
Lemma \ref{ln} i) and (\ref{N}). \bo.

\subsection{Incoming trajectory for small perturbations}

By Lemma 1.1, equation (\re{de})
with $y\in {\cal Y}$ is equivalent to
\begin{equation}\label{nu}
\Phi(y)=Rf,
\end{equation}
where
\begin{equation*}%\label{fi}
\Phi(y):=y-(R N)(y).
\end{equation*}
The map $RN: {\cal Y}\to {\cal Y}$ is continuous and  admits
the Fr\'echet differential $ (RN)'\in C({\cal Y},\mathscr{L}({\cal Y},{\cal Y}))$ by Lemma \ref{ln},
and
\begin{equation*}%\label{ps}
(RN)'=RN',~~~~(RN)'(0)=0.
\end{equation*}
Therefore, the map $\Phi$ is continuous ${\cal Y}\to {\cal Y}$,
$\Phi'\in C({\cal Y},{\cal{L}}({\cal Y},{\cal Y}))$, and $\Phi'(0)=I$, where $I$ is the identity operator.
\begin{theorem}\label{lnu}
Let $f\in L^2$. There exist $\varepsilon>0$, $C>0$ such that
equation (\ref{nu}) admits the unique solution $y\in {\cal Y}$ with $\|y\|< C$
for $\|f\|_{L^2}<\varepsilon$. This solution depends continuously on $f$.
\end{theorem}
\hspace{-5mm}{\bf Proof.} The map $\Phi: {\cal Y}\to {\cal Y}$ is continuously differentiable,
$\Phi(0)=0$  and $\Phi'(0)=I$.
Hence, by the Inverse Function Theorem
 (Theorem 10.4,
\cite{rr}) there exist $\varepsilon,~ C>0$
such that for $\|Rf\|<\varepsilon$ there exists the unique $y\in {\cal Y}$ with $\|y\|<C$ satisfying
(\ref{nu}) and depending continuously on  $Rf$.
It remains to note that   $R$ is  continuous
operator  $L^2 \to {\cal Y}$ by Lemma \ref{L1.1}. \bo

%%%%%%%%%%%%%%%%%%%%%%%%%%%%%%%%%%%%%%%%%%%%%%%%%%%%%%%%%%%%%%

\section{Asymptotic completeness}
\setcounter{equation}{0}
In this section we prove asymptotic completeness
for any hyperbolic stationary state.
First, we construct the incoming trajectory for large $t$ using Theorem \ref{lnu},
and afterwards we continue the trajectory backwards using a priory estimate.
\bt
 Let  conditions
(\re{L1.2b}) hold  and $F\in C^2.$
Then  system (\ref{L1.1}) is asymptotically complete at any hyperbolic stationary state.
\et
\hspace {-5mm}{\bf Proof.}
Let $S_+=(s_+,0)$ be a hyperbolic stationary state. We can consider $s_+=0$.
According to
Lemma \ref{tisp}, it suffices  to prove that
\medskip\\
i) for any $\Psi_+=(\Psi_0,\Psi_1)\in {\cal E}_{\infty}^+$ there exists a trajectory $y(t)$ satisfying
the differential equation (\ref{re2}) with $S$ defined by (\ref{S}) and  $y\to 0$, as $t\to \infty$.
\medskip
\\
ii) $\dot y\in L^2.$
\medskip
\\
First, let us decompose the function $F$ as
\begin{equation}\label{razF}
F(y)=Ay+N(y)
\end{equation}
where
$
A:=F'(0).
$
Then $A$ satisfies (\ref{A}) since $(0,0)$ is the hyperbolic state
and $N$ satisfies (\ref {N}), because $F \in C^2$.
Then equation (\ref{re2}) can be written as
\begin{equation}\label{ur}
\dot y = -\frac{1}{2}A-\frac{1}{2}N+f(t)
\end{equation}
where $f(t):=\frac{1}{2}{\dot S}\in L^2$ by Lemma \ref{ty0}.
Now we are able to prove i) and ii).
\medskip
\\
i) Let $T>0$ be such that $\|f\|_{L^2(T,\infty)}<\varepsilon$
for $t\geq T$, where $\varepsilon$ is chosen as in
Theorem \ref{lnu}.
Consider
 $$
 f_1(t):=\left \{\begin{array}{l}  f(t),~~ t\geq T\\
 0,~~~~t<T.
 \end{array}\right .
$$
By Theorem \ref{lnu} there exists $y_1(t)\in {\cal Y}$ satisfying equation (\ref{ur}).
Then $ y_1$ satisfies the inverse reduced equation (\ref{re2}) for $t\geq T$. It remains to
construct a solution $y_2$ to equation (\ref{ur}) or, equivalently, to (\ref{re2}) for
$0\leq t\leq T$ with the ``initial condition" $y_2(T)=y_1(T)$. It suffices to
 prove a priori estimate. Multiplying equation
(\ref{re2}) for $y_2$ by $2\dot y_2(t)$ and using (\ref{L1.2b}), we obtain that
\begin{equation*}%\label{am}
 (\nabla V)(y_2(t))\dot y_2(t)= 2|\dot{y_2}(t)|^2 - 2 f(t)\dot
y_2(t),~~~0< t < T.
\end{equation*}
Integrating and using the initial condition, we obtain
\begin{equation*}\label{am1}
V(y_1(T))-V(y_2(t))=2\int\limits_t^T|\dot y_2(\tau)|^2d\tau-2\int\limits_t^T f(\tau)\dot y_2(\tau)d\tau,~~~0\leq t\leq T.
\end{equation*}
Using the Young inequality, we estimate the second term
in the right hand side as
$$
2\left |\int\limits_t^T\dot S(\tau)\dot y_2(\tau)d\tau\right |\leq
\int\limits_t^T|\dot S(\tau)|^2d\tau+\int\limits_t^T|\dot
y_2(\tau)|^2d\tau.
$$
Hence,
\begin{equation}\la{am2}
V(y_2(t))+\int\limits_t^T|\dot y_2(\tau)|^2d\tau \leq
V(y_2(T))+\int\limits_t^T|f(\tau)|^2d\tau\leq B<\infty, ~~t\in [0,T]
\end{equation}
since $f\in L^2$.
Therefore, $y_2(t)$ is bounded for $t\in[0,T]$ by (\ref{L1.2b}).
Finally, defining
\begin{equation}\label{yott}
y(t):=\left \{\begin{array}{l}
 y_1(t),~~t\geq T,\\
 y_2(t), ~~t\in [0,T],
 \end{array}\right .
 \end{equation}
 we obtain that $y$ satisfies (\ref{re2}).
 Moreover, $y(t)\to 0$  by Def. (\ref{df}) and the fact that  $y_1\in {\cal Y}$.
 \medskip
 \\
 ii) Let us prove that $\dot y\in L^2$.
 First, $\dot y_1\in L^2$ since  $y_1$ satisfies (\ref{re2}),
 $\dot S\in L^2$,
 $$|F(y)|\leq C_\delta |y|~~ |y|<\delta$$
 by (\ref{ocn}) and (\ref{razF}).
Second,  the function $\dot y_2\in L^2$ by (\ref{am2}). So $\dot y\in L^2$ by (\ref{yott}).
 \bo
%%%%%%%%%%%%%%%%%%%%%%%%%%%%%%%%%%%%%%%%%%%%%%%%%%%%%%%%%%

%%%%%%%%%%%%%%%%%%%%%%%%%%%%%%%%%

\section{Counterexamples}
\setcounter{equation}{0}

In this section we give two examples which show that
 the incoming solution may not exist for nonhyperbolic stationary state. This means that the system is not asymptotically complete in this state.
\begin{example}
Let us consider equation (\ref{re2}) with $F$ satisfying (\ref{L1.2b}), and
$$
F(y)= 0,  ~~|y|<1.
$$
then $s_+=0$ is the nonhyperbolic stationary point.
Let us choose
\begin{equation}\label{f}
f(t)=\frac{1}{1+|t|}\in L^2.
\end{equation}
 Let us prove that
in this case a
trajectory satisfying condition
\be\la{y}
y(t)\to 0,~~t\to \infty.
\ee
does not exist.
In fact, let $y$ satisfy (\ref{re2}) with
$s_+=0.$
Then there exists $ T>0$
such that $|y(t)|<1/2$ for $t>T$.
Hence, we have $\dot y= 1/(1+t)$ for $t>T$, and
 therefore
$$y(t)=\ln(1+t)+C, ~~ t>T, $$
which contradicts (\ref{y}). \bo
\end{example}
\begin{example}
Let us consider equation (\ref{re2}) with $F$, satisfying (\ref{L1.2b}), and
$$
F(y)=y^2,  ~~|y|<1.
$$
Then $s_+=0$ is the nonhyperbolic stationary point.
Let us choose $f(t)$ from (\ref{f}) and prove that
the
trajectory satisfying  condition
(\ref{y})
does not exist.
In fact, let $y$ satisfy (\ref{re2})
with $s_+=0$.
Then there exists $ T>0$ such that $|y(t)|<1/2$ for $t>T$.
Hence,  $\dot y= y^2+1/(1+t)$ for $t>T,$ and therefore
$$
\dot y(t)\geq 1/(1+t),~~t>T.$$
Thus,
$$y(t)\geq \ln (1+t)+C,~~t>T $$ which contradicts (\ref{y}).\
\end{example}\bo

%%%%%%%%%%%%%%%%%%%%%%%%%%%%%%%%%%%%%%%

\end{document}